\documentclass[pra,twocolumn,showpacs]{revtex4}
\usepackage{graphicx}
\usepackage{psfrag}
\begin{document}

\title{Decoherence of molecular wave packets in an anharmonic potential}
\author{P\'{e}ter F\"{o}ldi}
\author{Mih\'{a}ly G. Benedict}
\email{benedict@physx.u-szeged.hu}
\author{Attila Czirj\'{a}k}
\author{Bal\'{a}zs Moln\'{a}r}

\affiliation{%
Department of Theoretical Physics, University of Szeged, 
Tisza L. k\"{o}r\'{u}t 84-86, H-6720 Szeged, Hungary}

\date{\today}
\begin{abstract}
The time evolution of anharmonic molecular wave packets
is investigated under the influence of
the environment consisting of harmonic oscillators.
These oscillators represent
photon or phonon modes and assumed to be in
thermal equilibrium.
Our model explicitly incorporates the fact that in the case of 
a nonequidistant spectrum the rates of the environment induced 
transitions 
are different for each transition.
The nonunitary time evolution is visualized by
the aid of the Wigner function related to the 
vibrational state of the molecule.
The time scale of decoherence is much
shorter than that of dissipation, 
and gives rise to states which are  mixtures of
localized states along the phase space orbit of the corresponding 
classical particle. This behavior is to a large extent independent 
of the coupling strength, the temperature of the environment
and also of the initial state. 
\end{abstract}
\maketitle

\section{introduction}
The correspondence between classical and quantum dynamics of anharmonic 
systems has gained significant attention in the
past few years \cite{PS86,AP89,VE94,AS00}.
A short laser pulse impinging on an atom or a molecule excites 
a superposition of several stationary states, and the resulting 
wave packet follows the orbit of the corresponding classical 
particle in the initial stage of the time evolution.

However, the nonequidistant nature of the involved energy 
spectra causes peculiar quantum effects, broadening of the initially
well localized wave packets, revivals and partial revivals 
\cite{AP89,PS86,VE94,AS00,LAS96,DK00}. Partial revivals are in close connection
with the formation of Schr\"odinger-cat states, which, in this context, 
are coherent superpositions of two spatially separated, 
well localized wave packets \cite{JV94}. 
Phase space description of vibrational 
Schr\"odinger-cat state formation using animated Wigner
functions can be found in \cite{FC02}.
However, these highly nonclassical states are expected to be
particularly sensitive to decoherence \cite{GJK96}. The aim of this
paper is to analyze the process of decoherence for the spontaneously
formed Schr\"odinger-cat states in the anharmonic potential.

We consider the decoherence model
which relies on the interaction of the quantum system 
(\textsl{S}) with its environment (\textsl{E}) that has many 
degrees of freedom.
The dynamics of the environment together with the quantum system
under investigation 
is unitary, i.e., the density operator $\rho_{SE}$ 
of the coupled $S+E$ system always represents a pure state.
Starting from 
an initially uncorrelated density operator, 
the interaction builds up \textsl{S}--\textsl{E} entanglement and  
the reduced density operator 
\begin{equation}
\rho_S=\mathrm{Tr}_E \left( \rho_{SE} \right),
\end{equation} 
where $\mathrm{Tr}_E$ means a trace over the environmental degrees of freedom, 
turns into a mixture. 
Assuming negligibly short relaxation times in the environment,
one obtains a Markovian master equation  
which is a useful tool for the dynamical investigation of the 
environment induced decoherence.
It allows for the determination of pointer states \cite{Z81}, 
that is, states that are 
favored by decoherence \cite{ZHP93,FCB01a}, and also for the
calculation of the characteristic time of decoherence
for different initial states \cite{FCB01a,BC99}.

In the following we introduce 
a master equation that takes into account the fact that in a general
anharmonic system the relaxation rate of each energy eigenstate is different.
This master equation is applied to the case of wave packet motion in 
the Morse potential that is often used to describe a 
vibrating diatomic molecule. 
Considering the  phase-space description of decoherence,
we show how the phase portrait of the system reflects the damping of 
revivals in the expectation values of the position and momentum operators due
to the effect of the environment. 
We also calculate and plot 
the time evolution of the Wigner function corresponding
to the reduced density operator of the Morse system. 
The Wigner function picture visualizes the fact that although
our master equation reduces to the 
amplitude damping equation \cite{WM85,SW85,P90,BBK92} in 
the harmonic limit,
the anharmonic
effects lead to a decoherence scheme which is similar to the
phase relaxation \cite{WM85} of the harmonic oscillator (HO).
It is found that the reduced density operator that
arises due to the decoherence 
can be identified with a mixture of states that are well-localized
in the phase space and equally distributed along the orbit of the 
corresponding classical particle. We illustrate the generality of this
decoherence scheme by presenting the time evolution of an energy
eigenstate as well.  
We also calculate the decoherence time for various 
initial wave packets. We show that decoherence is faster for wave packets
that correspond to a classical particle with a phase space
orbit of larger diameter.

\section{Description of the model}
\label{mastersec}
The total Hamiltonian of an anharmonic
system and its environment has the form
\begin{equation}
H_{S+E}=H_S+H_E+V.
\label{MEham}
\end{equation}
The self-Hamiltonian of the system is written as:
\begin{equation}
H_S=\sum_n E_n |\phi_n\rangle\langle\phi_n|,
\label{MEsystemH}
\end{equation}
where the spectrum $\{E_n\}$ is assumed to be nondegenerate 
and discrete, but not
necessarily equidistant. $E_0$ denotes the ground state energy of the system,
and $E_m>E_n$ whenever $m>n$. 
The environment is represented by a set of harmonic oscillators
\begin{equation}
H_E=\sum_k \hbar\omega_k (a^{\dagger}_k a_k+1/2).
\label{MEenvH}
\end{equation}
We assume the following interaction Hamiltonian
\begin{equation}
V=\hbar\mathcal{X}^{\dagger}\sum_k g_k a_k+
\hbar\mathcal{X}\sum_k g_k a^{\dagger}_k,
\label{MEinterH}
\end{equation}
where $\mathcal{X}$ is an appropriate system operator 
that transforms each
eigenstate of $H_S$ into a superposition of different eigenstates 
corresponding to \emph{lower} energy values. 
This is the application of the rotating wave approximation (RWA)
to an anharmonic, multilevel system.
The operator $\mathcal{X}^{\dagger}$ 
is the Hermitian conjugate of $\mathcal{X}$, and for the sake of simplicity 
the coupling constants $g_k$ were taken to be real.
In the case of a vibrating diatomic molecule $H_E$ describes the 
modes of the free electromagnetic field, while the
terms $\mathcal{X}$ and $\mathcal{X}^\dagger$ in the interaction Hamiltonian 
are related to the molecular dipole moment operator.

We assume that the environment is in thermal equilibrium 
at a given temperature $T$, and the 
environmental oscillators have continuous distribution
with frequency dependent density of states, $D(\omega)$. 
Starting from the von Neumann equation
for the total density operator, standard techniques \cite{N58,Z60a,H73,WM94}
lead to the following Markovian master equation in the Schr\"{o}dinger picture 
\begin{eqnarray}
\frac{d}{dt}\rho_{S}&=&-{\frac{i}{\hbar}}\left[H_S,\rho_{S}\right]
-\big(
\mathcal{X}^{\dagger}\mathcal{X}_{e}\rho_{S}
+\mathcal{X}\mathcal{X}_{a}^{\dagger}\rho_{S}\nonumber 
\\
&-&\mathcal{X}_{a}^{\dagger}\rho_{S}\mathcal{X}
-\mathcal{X}_{e}\rho_{S}\mathcal{X}^{\dagger} + \mathrm{h.c.}
\big).
\label{ME}
\end{eqnarray}
The matrix elements of the operators appearing in the nonunitary terms 
are given by
\begin{eqnarray}
\langle \phi_m |\mathcal{X}_{a}|\phi_n\rangle&=&\pi
\langle \phi_m |\mathcal{X}|\phi_n\rangle \ D(\omega_{n m}) g^2(\omega_{n m})
\overline{n}(\omega_{n m}), \nonumber\\
\langle \phi_m |\mathcal{X}_{e}|\phi_n\rangle&=&\pi
\langle \phi_m |\mathcal{X}|\phi_n\rangle \ 
D(\omega_{n m}) g^2(\omega_{n m}) \nonumber \\
&\times&\left(\overline{n}(\omega_{n m})+1\right),
\label{MEXelements}
\end{eqnarray}
where $\omega_{n m}=|E_m-E_n|/\hbar$, 
$\overline{n}(\omega)=({\exp{\frac{\hbar\omega}{kT}}-1})^{-1}$
denotes the average number of quanta in the corresponding mode of the 
environment, and the subscripts $e$ and $a$ refer to emission and absorption,
respectively.

As we can see, the matrix elements (\ref{MEXelements}) of the operators 
that induce the transitions 
depend on the Bohr frequency of the involved transition,
which is a genuine anharmonic feature.  
In the special case of the HO, when
$H_S$ has equidistant spectrum, and
$\mathcal{X}$ is identified with the usual annihilation operator $a$, 
both $\mathcal{X}_{a}$ and $\mathcal{X}_{e}$ are
proportional to $\mathcal{X}\equiv a$, and Eq.~(\ref{ME}) reduces to
the amplitude damping master equation \cite{WM85,SW85,P90,BBK92} 
at a finite temperature.

In certain cases one can further simplify Eq.~(\ref{ME}). When
the environment induced relaxation rates are much lower than the relevant
Bohr frequencies, the system Hamiltonian induces oscillations
that are very fast even on the time scale of decoherence and vanish
on the average. Ignoring these fast oscillations we arrive at the
interaction picture master equation 
\begin{equation}
\frac{d}{dt}\langle\phi_i|\rho_{S}|\phi_j\rangle=
\delta_{i,j}\sum_{k\neq i}\gamma_{ik}\langle\phi_k|\rho_{S}|\phi_k\rangle
-\Gamma_{ji}^c \langle\phi_i|\rho_{S}|\phi_j\rangle,
\label{ME2}
\end{equation}
that has already been obtained in Refs.~\cite{L66, A73}
in order to treat the spontaneous emission
of a multilevel atom. 
In Eq.~(\ref{ME2}), $\gamma_{ik}$ denotes a relaxation rate, that is 
the probability of the $|\phi_k\rangle\rightarrow|\phi_i\rangle$ 
transition per unit time, while $\Gamma_{ji}^c=1/2\sum_k(\gamma_{ik}+
\gamma_{jk})$, where
\begin{equation}
\gamma_{ik}=\left\{
\begin{array}{lr}
2\  \langle \phi_i |\mathcal{X}_{e}|\phi_k\rangle
\langle \phi_i |\mathcal{X}|\phi_k\rangle
&{\textnormal{if}}\  i<k,
\\
0 & {\textnormal{if}}\  i=k,
\\
2\  \langle \phi_i |\mathcal{X}_{a}|\phi_k\rangle
\langle \phi_i |\mathcal{X}|\phi_k\rangle
& {\textnormal{if}}\  i>k.
\label{MEpoptrprob}
\end{array}
\right.
\end{equation}

However, due to the elimination of 
the fast oscillations related to $H_S$, Eq.~(\ref{ME2}) is not suitable 
for investigating the wave packet motion and decoherence simultaneously,
therefore we propose to use Eq.~(\ref{ME}). On the other hand we note that 
Eq.~(\ref{ME2}) radically reduces the computational costs of calculating
the time evolution for long times, which might be necessary when
the system-environment coupling is very weak.

Supposing that our knowledge is limited to
the populations $P_n=\langle \phi_n|\rho_s|\phi_n\rangle$,
both Eq.~(\ref{ME}) and Eq.~(\ref{ME2}) leads to the Pauli type equation
\begin{equation}
\frac{d}{dt}P_n=\sum_k \left( \gamma_{nk} P_k-\gamma_{kn}P_n \right).
\label{pauli}
\end{equation}
Requiring the condition of detailed 
balance \cite{R65} in Eq.~(\ref{pauli}) leads to the steady-state 
thermal distribution at the temperature of the environment.
\vskip 12pt

When a diatomic molecule is considered in the environment of the
free electromagnetic field, 
the operators $\mathcal{X}$ and $\mathcal{X}^\dagger$
in Eq.~(\ref{MEinterH})
gain a clear interpretation: in the eigenbasis of $H_S$ 
they are the upper and lower 
triangular parts of the molecular dipole moment operator, $\hat\mu$.
We will assume that $\hat\mu$ is linear \cite{YL98}, that is, proportional 
to the displacement $X$ of the center of mass of the diatomic system
from the equilibrium position.
In this case $g^2(\omega)D(\omega)\propto \omega^3$,
that is,
\begin{eqnarray}
\langle \phi_m |\mathcal{X}_{a}|\phi_n\rangle&=&\lambda\ 
\langle \phi_m |X|\phi_n\rangle \ {\omega_{n m}}^3\ 
\overline{n}(\omega_{n m}), \nonumber  \\
\langle \phi_m |\mathcal{X}_{e}|\phi_n\rangle&=&\lambda\ 
\langle \phi_m |X|\phi_n\rangle \  {\omega_{n m}}^3 
\left(\overline{n}(\omega_{n m})+1\right),
\label{MorseX}
\end{eqnarray}
where matrix elements of $X$ can be calculated using 
the algebraic method summarized in \cite{BM99}, and
$\lambda=\pi g^2(\omega)D(\omega)/\omega^3$
is an overall, frequency independent coupling constant. 

However, in order to get insight into the 
interplay between wave packet motion and decoherence, it
is worth considering a stronger molecule-environment interaction
than the electromagnetic field modes can provide.
Keeping the structure of Eqs.~(\ref{MorseX}), this can be done by
increasing the value of $\lambda$.
Here we present calculations with two different
coupling constants, $\lambda_1$  and  $\lambda_2$ which are chosen 
so that at zero temperature  
$\omega_{01}/\gamma_{01}\approx10^5$ and $4\times 10^3$ 
for $\lambda=\lambda_1$ and $\lambda_2$, respectively. 
This model allows for the numerical integration of  
the master equation (\ref{ME}) (that provides more details of the 
dynamics than Eq.~(\ref{ME2}))
in a time interval that is long enough to identify the
effects of decoherence. These effects can be summarized
in a decoherence scheme (see Sec.~\ref{wignersec}) that has 
a clear physical interpretation, and which is valid also in the weak
molecule-environment interaction, when (\ref{ME2}) is more efficient to
calculate the time evolution.

\vskip 12pt 
In the following we shall apply our master equation (\ref{ME}) 
to the case of the Morse Hamiltonian \cite{HH79} that is often used to
describe a vibrating diatomic molecule.
This Hamiltonian has the dimensionless form
\begin{equation}
H_S=P^2+(s+1/2)^2[\exp(-2X) -2\exp(-X)],
\label{ham}
\end{equation}
where the shape parameter $s$ is related to the dissociation
energy $D_{\mathrm{diss}}$, the reduced mass of the molecule $m$, 
and the range parameter of
the potential $\alpha$ via 
$s=\frac{\sqrt{2mD_{\mathrm{diss}}}}{\hbar \alpha}-1/2.$
The dimensionless displacement and momentum operators 
obey the canonical commutation relation $[X,P]=i$.  
 
The Hamiltonian (\ref {ham}) sustains
$[s]+1$ normalizable eigenstates (bound states),
corresponding to the eigenvalues 
$E_{m}(s)=-(s-m)^{2},$ $m=0,1,\ldots \lbrack s]$,
where $[s]$ denotes the largest integer that is 
smaller than $s$. 
The continuum above $E_{m=[s]}$ corresponds to the dissociated molecule
with positive energies. For the sake of definiteness we have
chosen the NO molecule as our model, where $s=54.54$.

The initial wave packets of our analysis will be Morse coherent 
states \cite{BM99}, $|x_0,p_0\rangle$,
which are localized on 
the phase space around the point $(x_0, p_0)$.
The Wigner function of a representative Morse coherent state
is shown in Fig.~\ref{wig1} a).
These states
can be prepared by an appropriate
electromagnetic pulse that drives the vibrational state of the molecule
starting from the ground state into an approximate  coherent state. 
An example can be found in \cite{MBF01}, where the effect of an 
external sinusoidal field is considered.

Let us note that the construction given in \cite{MF02}
would allow us to use arbitrary initial states, 
but for our current purpose it suffices to consider
states $|x_0,p_0\rangle$ with negligible
dissociation probability, i.e., coherent states that practically
can be expanded in terms of the bound states
$|\phi_n\rangle$,  $n=0,1,\ldots \lbrack s]$.

\section{Time evolution of the expectation values}
\label{expectsec}
Starting from  $|\psi(t=0)\rangle=|x_0,p_0=0\rangle$ as initial states,
the qualitative behavior of the expectation value 
$\langle X\rangle(t)=\langle\psi(t)| X|\psi(t)\rangle$ draws the limit
of \textit{small} oscillations. In the absence of environmental
coupling (i.e., $\lambda=0$), for $x_0\leq0.05$, $\langle X\rangle(t)$
(as well as $\langle P\rangle(t)$)
exhibits sinusoidal oscillations. 
\begin{figure}[htb]
\begin{center}
\psfrag{plabel}[tl][tl]{$\langle P \rangle$}
\psfrag{xlabel}{$\langle X \rangle$}
\psfrag{tlabel}{$t/t_0$}
\includegraphics*[bb=80 50 670 470 , width=8.5cm]{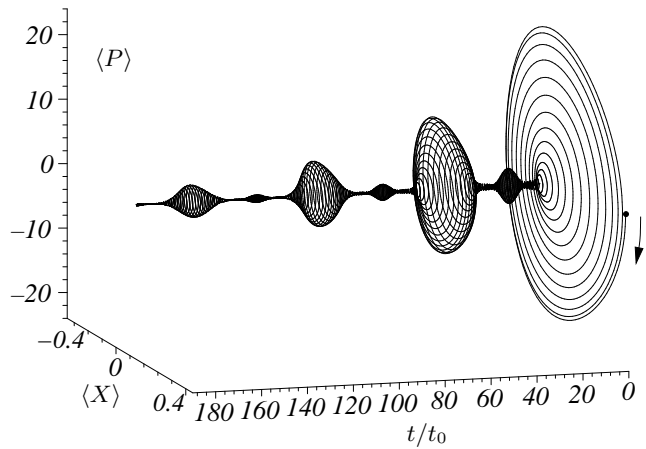}
\caption
{
Phase portrait corresponding to the time evolution of the initial state
$|\psi(t=0)\rangle=|x_0,0\rangle$, with $x_0=0.5$. 
The parameters are $\lambda=\lambda_1$, $T=5 \ \hbar\omega_{01}/k$, and
$t_0$ is the period of the small oscillations in the 
potential. The initial point $\langle X \rangle=0.5$, $\langle P \rangle=0$
together with the starting direction is also indicated.  
\label{portrait}
}
\end{center}
\end{figure}
For larger initial displacements from
the equilibrium position, the anharmonic effects become apparent: 
the amplitude of the oscillations
decreases almost to zero, then faster oscillations with
small amplitude appear but later we re-obtain almost exactly 
$\langle X\rangle(0)$ (and $\langle P\rangle(0)$ as well), and the whole
process starts again \cite{FC02}.  
Without the influence of the environment
the main attributes of $\langle X\rangle(t)$ and $\langle P\rangle(t)$ 
can be explained by referring
to the various Bohr frequencies that determine their time dependence:
dephasing of these frequencies leads to the collapse of the expectation 
value, and we observe revival when they rephase again. 

For the initial state of $|\psi(t=0)\rangle=|x_0,0\rangle$, with $x_0=0.5$,
the original phase of the 
eigenstates is restored \cite{PS86,AP89} 
around the full revival time $t_{rev}=110\ t_0$, where $t_0$ is the period 
of the small oscillations in the potential.   
At $t/t_0=55$ and $t/t_0=27.5$
half and quarter revivals \cite{PS86,AP89} can be observed.
Fig.~\ref{portrait} shows the damping of the revivals both in
$\langle X\rangle(t)$ and $\langle P\rangle(t)$
when interaction with the environment is turned on. 
Note that the phase portrait of the corresponding classical particle
would be a helix with monotonically decreasing diameter,
revivals are of quantum nature. However, Fig.~\ref{portrait}
does not provide a complete description of the time evolution in
the phase space, this can be given by
using Wigner functions, see Sec. \ref{wignersec}.

\section{Decoherence times}
\label{dectimesec}
Our master equation (\ref{ME}) describes decoherence as well
as dissipation. However, the time scale of these processes is generally
very different, and one can distinguish the stages of the
time evolution that are dominated either by decoherence or 
dissipation \cite{FCB01a}.
\begin{figure}[htb]
\begin{center}
\psfrag{elabel}[tl][tl][0.8]{$S(t)$}
\psfrag{trlabel}[bl][bl][0.8]{$\mathrm{Tr}\left[\rho_S^2(t)\right]$}
\psfrag{eaxis}{$S$}
\psfrag{traxis}[br][bl]{$\mathrm{Tr}\rho_S^2$}
\psfrag{taxis}{$t/t_0$}
\includegraphics*[bb=65 45 775 515 , width=8.5cm]{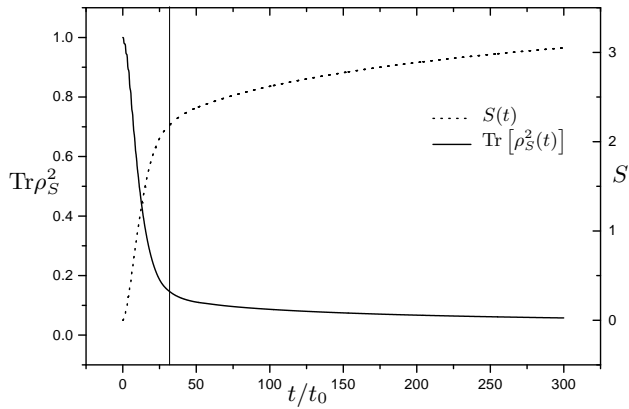}
\caption
{
The entropy and the purity of the reduced density matrix of the Morse system
as a function of time, calculated using Eq.~(\ref{ME}). The coupling parameter
(see Sec.~\ref{mastersec}) is $\lambda=\lambda_1$ and
$T=10 \ \hbar\omega_{01}/k$. The initial state was  
$|\psi(t=0)\rangle=|x_0,0\rangle$, with $x_0=2.0$.
\label{dtdef}
}
\end{center}
\end{figure}
In Fig.~\ref{dtdef} an example is depicted showing how the method of 
time scale separation works.
We have calculated the entropy 
\begin{equation}
S=-\mathrm{Tr}\left[\rho_S \ln(\rho_S)\right],
\end{equation}
as well as the quantity $\mathrm{Tr}[\rho_S^2]$, 
which measures the purity of the reduced density operator. Note that
the $\mathrm{Tr}$ operation without subscript refers to the
trace in the system's Hilbert space.
Decoherence time $t_d$ is defined as the
time instant 
that divides the time axis into two parts
where the character of the physical process is clearly different. 
Initially both $S(t)$ and $\mathrm{Tr}[\rho_S^2(t)]$ change rapidly 
but having passed  $t_d$ (emphasized by a vertical line in 
Fig.~\ref{dtdef}), the moduli of their derivative
significantly decrease. After $t_d$ the entropy and the purity 
change on the time scale which is characteristic of 
the dissipation of the system's energy during the whole process
\cite{BC99,FCB01a}. 
In other words, decoherence dominated time evolution
turns into dissipation dominated dynamics around $t_d$.
In the next section we shall determine the density operators 
into which the process of decoherence drives the system. 
In connection with these results we have verified that 
the states around the decoherence time 
do not change appreciably in a time interval
that covers the possible errors in determining $t_d$.

An interesting question is the dependence of the decoherence time on the 
initial state of the time evolution. 
We calculated $t_d$ as a function of the initial displacement
for the case of displaced 
ground states (that is, coherent states with zero momentum, 
$|x_0,0\rangle$) as initial states.
It was found that for all values of $\lambda$ and $T$, the
decoherence time is longer for smaller initial displacements. 
Additionally, for fixed $\lambda$ and $T$ the function
$t_d(x_0)$ can be well approximated by an exponential curve
$t_d(x_0)=t_d(0) \exp(-\kappa x_0)$. E. g., for $\lambda=\lambda_1$,
$T=10 \ \hbar\omega_{01}/k$ and $0<x_0\leq 2$ the parameters take 
the values $t_d(0)=93\  t_0$ and $\kappa=0.97$.

It is known \cite{AP89} that quarter revivals in an anharmonic potential
lead to the formation of Schr\"{o}dinger-cat states, i.e., states that are
superpositions of two distinct states localized in space 
\cite{AP89} as well as in momentum \cite{FC02,KP95}. 
On the other hand,   
smaller initial displacements correspond to classical phase space orbits
with smaller diameter. Consequently the quantum interference related to 
nonclassical states that are formed during the course of time 
cover a smaller area in the phase-space in this case. 
This means that our result is a manifestation of the general
feature of decoherence that increasing the ``parameter of nonclassicality'',
which is the diameter of the corresponding classical orbit in our case,
causes faster decoherence \cite{GJK96}. 
A similar result was found in \cite{FCB01a} for the case of 
decoherence in a system of two-level atoms \cite{BC99,BBH00}.

\section{Wigner function description of the decoherence}
\label{wignersec}
\begin{figure*}[htp]
\begin{center}
\includegraphics*[bb=60 60 1370 1070 , width=17.5cm]{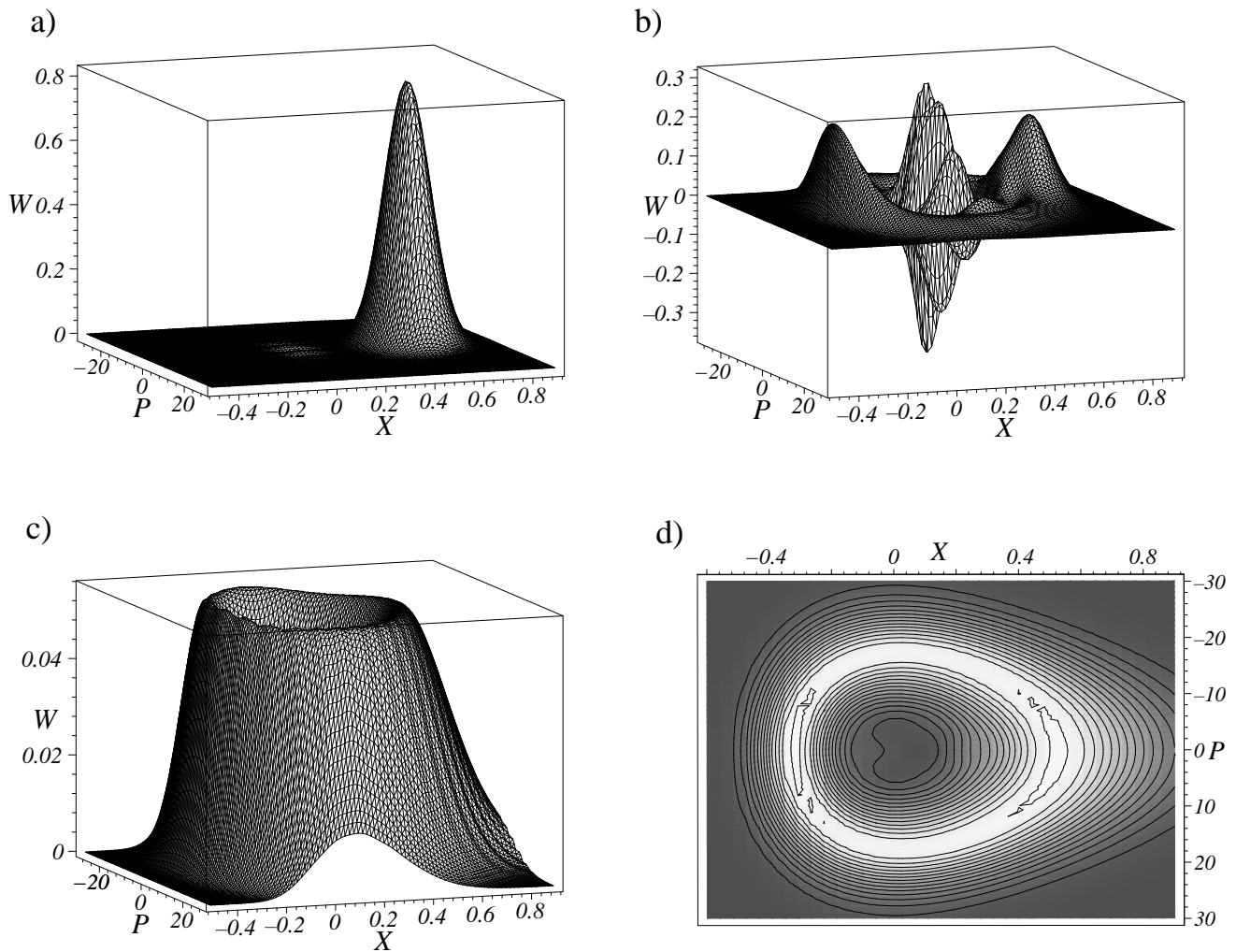}
\caption
{
Time evolution of the Wigner function corresponding to the initial state
$|\psi(t=0)\rangle=|x_0,0\rangle$, with $x_0=0.5$. The coupling parameter is
(see Sec.~\ref{mastersec}) $\lambda=\lambda_2$ and
$T=0.3 \ \hbar\omega_{01}/k$. The plots a) and b) 
correspond to time instants $t/t_0=0\  \text{and}\  27.5$, while
both c) and the contour plot d) are snapshots taken at $t/t_0=137.5$.  
\label{wig1}
}
\end{center}
\end{figure*}
In order to visualize the time evolution of the reduced density matrix of the
Morse system we have chosen the Wigner function picture, which represents
$\rho_S$ as a function over the classical phase space
\begin{equation}
W(x,p,t)={\frac{1}{2\pi}}\int_{-\infty}^{\infty} \langle x-u/2|
\rho_S(t)|x+u/2\rangle e^{i up} d u.
\end{equation} 
This description 
allows us to investigate the correspondence between classical and quantum 
dynamics.

First we consider the ideal case without environment.
Then, in the initial stage of the time evolution, the positive hill 
corresponding to the wave packet $|x_0,p_0\rangle$
follows the orbit of the classical particle that has started 
from $(x_0,p_0)$ at $t=0$. However, due to the uncertainty relation,
the Wigner function as a quasiprobability distribution has a finite
width, and this fact combined with the form of the Morse potential 
implies the stretching of the Wigner function along the classical 
orbit in the course of time. (See Ref.~\cite{KP95} for similar results
with the Husimi $Q$ function.) After a certain time the
increasingly broadened wave packet
becomes able to interfere with itself,
and around the quarter revival time one can observe two positive hills
chasing each other at the opposite sides of the classical orbit.
The strong oscillations of $W$ between the hills represent the quantum 
correlation of the constituents of this molecular Schr\"{o}dinger-cat 
state \cite{JV94}.
Later on the initial Wigner function is restored almost exactly and
Schr\"{o}dinger-cat state formation starts again. Detailed 
Wigner function description of
these processes that are related to the free time evolution can be found
in \cite{FC02}. 

In the case when environmental effects are present,
we found that decoherence follows a general scheme. A representative series 
of Wigner functions is shown in Fig.~\ref{wig1}. The snapshots correspond
to the initial state and time instants when the first and third
Schr\"{o}dinger-cat state formation  would occur in the absence of 
the environment. Consequently, the Wigner function in Fig.~\ref{wig1} b) 
corresponds almost to a Schr\"{o}dinger-cat state, but this state is 
already a mixture. However, there are still negative parts of the function
in between the positive 
hills centered at $x_1=0.51,\  p_1=0$ and  $x_2=-0.34,\  p_2=0$.
The ``ridge'' that connects these hills
along the classical orbit is absent in a pure Schr\"{o}dinger-cat state. 
Later on this 
ridge becomes more and more pronounced and at the decoherence time 
we arrive at the positive Wigner
function of  Fig.~\ref{wig1} c) and d). According to the contour
plot  Fig.~\ref{wig1} d), the highest values of this function trace out
the phase space orbit of the corresponding classical particle.
That is, $\rho_S^{dec}$, the reduced density matrix 
that arises as a result of decoherence, 
can be interpreted as a mixture of localized states that are
equally distributed along the
orbit of the corresponding classical phase space orbit. 
\begin{figure*}[htb]
\begin{center}
\includegraphics*[bb=50 60 1350 1100 , width=17.5cm]{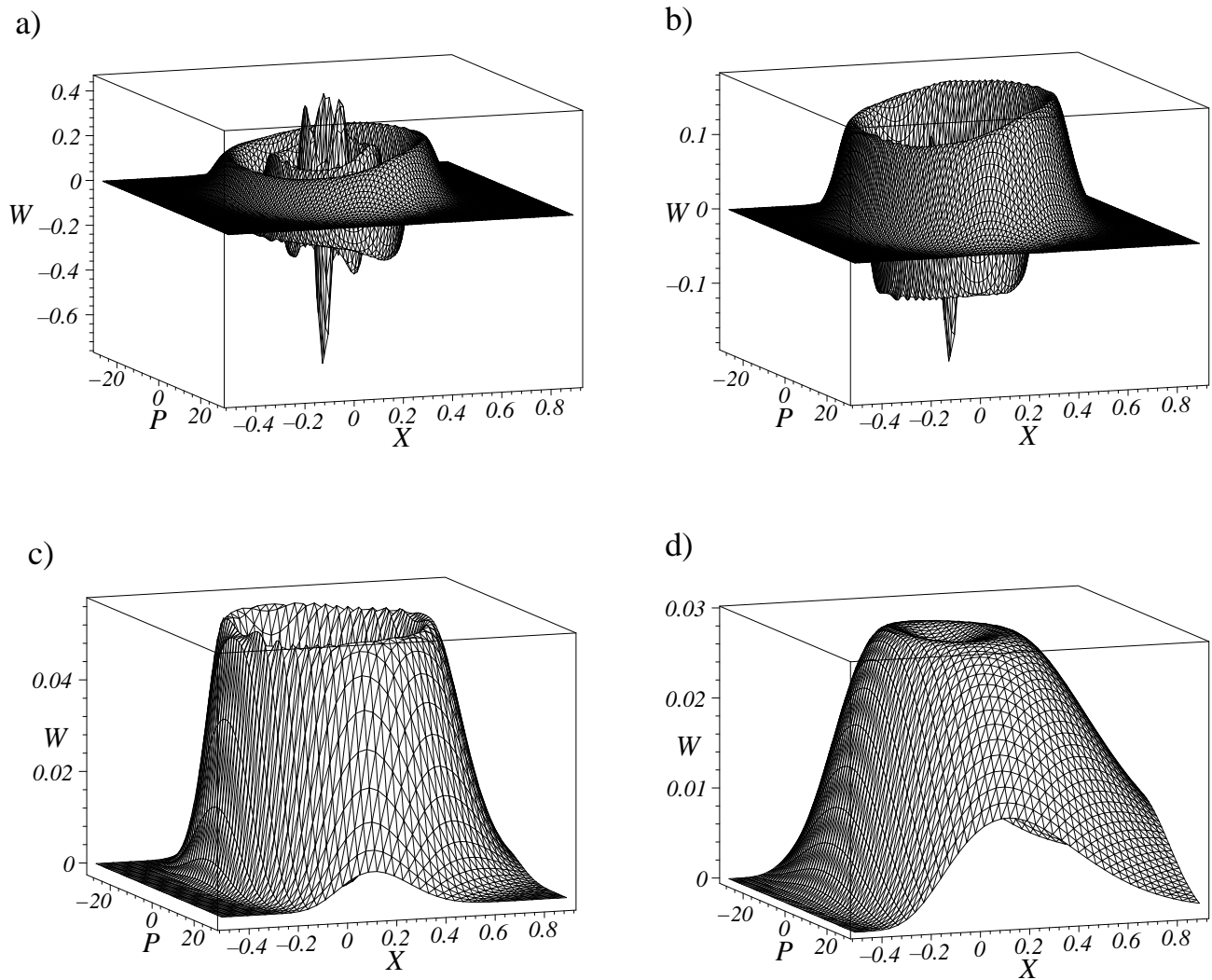}
\caption
{
Time evolution of the Wigner function corresponding to the fifth bound
state as initial state.
The coupling parameter
(see Sec.~\ref{mastersec}) is  $\lambda=\lambda_1$ and
$T=10 \ \hbar\omega_{01}/k$. The plots a), b), c) and d)
correspond to time instants $t/t_0=0,\   27.5, \ 330$  and $1000$, 
respectively.
\label{wig2}
}
\end{center}
\end{figure*}

It is worth comparing this result with the case of the HO, when 
the master equation (\ref{ME}) reduces to the
amplitude damping equation \cite{WM85,SW85,P90,BBK92}, 
see Sec.~\ref{mastersec}.
It is known that harmonic oscillator coherent states are
robust against the decoherence described by the 
amplitude damping master equation (as well as against the Caldeira-Leggett
\cite{CL83} master equation \cite{ZHP93}), 
the initial superposition of coherent 
states turns into the statistical mixture of essentially the
same states. This is a consequence of the facts that these
states are eigenstates of the destruction operator $a$, and
the operators in the nonunitary terms of Eq.~(\ref{ME}) 
are proportional to $a$ and $a^{\dagger}$ in the harmonic case. 
None of these statements can be transferred to the anharmonic system, 
where the Morse coherent states do not remain localized during the
course of time, even without environment.
Therefore the scheme of the decoherence is qualitatively different for the
harmonic and anharmonic oscillators: Our results in the anharmonic system are
similar to the phase relaxation in the harmonic case \cite{WM85}, where the
energy of the system remains unchanged, but the phase information is 
completely destroyed. 
We note that a similar result was obtained in Ref.~\cite{B01},
where the rotational degrees of freedom were considered as a 
reservoir for the harmonic vibration of hot alkaline dimers.

Our decoherence scheme is universal to a large extent.
In the investigated domain of the coupling constants $\lambda_1\leq \lambda
\leq\lambda_2$ and temperatures ranging from $T=0$ 
to $T=15 \ \hbar\omega_{01}/k$,
it is found to be valid for all initial states, not only for coherent states.
Fig.~\ref{wig2} shows an example when the initial state is not a wave packet,
it is the fifth bound state, corresponding to $E_5$, which
is very close to $\langle0.5,0|H_S|0.5,0\rangle$, so direct comparison
with Fig.~\ref{wig1} is possible. As we can see, although the two Wigner
functions are initially obviously very different, they follow different routes
(that takes different times) to the \textit{same} 
state: Fig.~\ref{wig1} c) and 
Fig.~\ref{wig2} c) are practically identical. The final plot 
in Fig.~\ref{wig2} indicates how the Wigner function represents the 
long way to thermal equilibrium with the environment: the distribution
becomes wider and the hole in the middle disappears.    

It is expected that the loss of phase information has observable
consequences. According to the Franck-Condon principle,
the absorption spectrum of a molecule around
the frequency corresponding to an electronic transition between 
two electronic surfaces depends on the vibrational state. 
The time dependence of the spectrum should exhibit the 
differences between the pure state of an oscillating wave packet and
the state $\rho_S^{dec}$ and the thermal state.
More sophisticated experimental methods based on the
detection of fluorescence \cite{KZ90} or fluorescence
intensity fluctuations \cite{WT00}, surely have the capacity
of observing the dephasing phenomenon considered in this paper.

\section{Conclusions}
\label{conclusionsec}
We investigated the decoherence of wave packets in the Morse potential.
The decoherence time for various initial states was calculated and it
was found that the larger is the diameter of the phase space orbit described 
by a wave packet, the faster is the decoherence. We obtained a general
decoherence scheme, which has a clear physical interpretation: The reduced
density operator that is the result of the decoherence is a mixture of
states localized along the corresponding classical phase space orbit.

This work was
supported by the Hungarian Scientific Research Fund (OTKA) under contracts
Nos. T32920, D38267, and by the Hungarian Ministry of Education under contract 
No. FKFP 099/2001.

\end{document}